\documentclass[aps,superscriptaddress,floats,showpacs,floatfix,onecolumn]{revtex4}
\usepackage{bm}
\usepackage{graphicx}
\usepackage{subfigure}
\usepackage{amssymb}
\usepackage{amsmath}
\usepackage{bbm}
\usepackage{color}
\usepackage{hyperref}
\hypersetup{colorlinks, citecolor = blue, filecolor = blue, linkcolor = blue, urlcolor = blue}
\usepackage[a4paper,left=1.5cm, right=1.3cm, top=2cm,bottom=2cm]{geometry}
\begin{document}
\newcommand{\neu}[1]{\textcolor{blue}{#1}}
\newcommand{\joerg}[1]{\textcolor{red}{#1}}
\newcommand{\kathrin}[1]{\textcolor{blue}{#1}}
\newcommand{\christiane}[1]{\textcolor{green}{#1}}

\title{Probing turbulent superstructures in Rayleigh-B\'{e}nard convection \\by Lagrangian trajectory clusters}
\author{Christiane Schneide}
\affiliation{Institut f\"ur Mathematik und ihre Didaktik, Leuphana Universit\"at L\"uneburg, D-21335 L\"uneburg, Germany}
\author{Ambrish Pandey}
\affiliation{Institut f\"ur Thermo- und Fluiddynamik, Technische Universit\"at Ilmenau, Postfach 100565, D-98684 Ilmenau, Germany}
\author{Kathrin Padberg-Gehle}
\affiliation{Institut f\"ur Mathematik und ihre Didaktik, Leuphana Universit\"at L\"uneburg, D-21335 L\"uneburg, Germany}
\author{J\"org Schumacher}
\affiliation{Institut f\"ur Thermo- und Fluiddynamik, Technische Universit\"at Ilmenau, Postfach 100565, D-98684 Ilmenau, Germany}
\affiliation{Tandon School of Engineering, New York University, New York, NY 11201, USA}
\date{\today}

\begin{abstract}
We analyze large-scale patterns in three-dimensional turbulent convection in a horizontally extended square 
convection cell by Lagrangian particle trajectories calculated in direct numerical simulations. A simulation 
run at a Prandtl number Pr $=0.7$, a Rayleigh number Ra $=10^5$, and an aspect ratio $\Gamma=16$ is therefore 
considered. These large-scale structures, which are denoted as turbulent superstructures of convection, are 
detected by the spectrum of the graph Laplacian matrix. Our investigation, which follows Hadjighasem {\it et al.}, Phys. 
Rev. E {\bf 93}, 063107 (2016), builds a weighted and undirected graph from the trajectory points of Lagrangian 
particles. Weights at the edges of the graph are determined by a mean dynamical distance between different particle trajectories. 
It is demonstrated that the resulting trajectory clusters, which are obtained by a subsequent $k$-means clustering, coincide with 
the superstructures in the Eulerian frame of reference. Furthermore, 
the characteristic times $\tau^L$ and lengths $\lambda_U^L$ of the superstructures in the Lagrangian frame of 
reference agree very well with their Eulerian counterparts, $\tau$ and $\lambda_U$, respectively.  This 
trajectory-based clustering is found to work for times $t\lesssim \tau\approx\tau^L$. Longer time periods $t\gtrsim \tau^L$ 
require a change of the analysis method to a density-based trajectory clustering by means of time-averaged Lagrangian 
pseudo-trajectories, which is applied in this context for the first time.  A small coherent subset of the pseudo-trajectories is obtained in this way consisting of those 
Lagrangian particles that are trapped for long times in the core of the superstructure circulation rolls and are thus 
not subject to ongoing turbulent dispersion. 
\end{abstract}
\keywords{}
\maketitle

\section{Introduction}
Compared to investigations of turbulent Rayleigh-B\'{e}nard convection in the Eulerian frame of reference \cite{Kadanoff2001,Ahlers2009,Chilla2012}, 
the number of studies in the Lagrangian perspective in which the turbulent fields and the 
related transport are monitored along the trajectories of massless point-like particles \cite{Toschi2009}
has remained surprisingly small to date. Numerical studies were primarily 
focused on the turbulent dispersion of Lagrangian particle pairs \cite{Schumacher2008} and tetrads \cite{Schumacher2009}, the entropy production 
rates along individual Lagrangian trajectories \cite{Zonta2016} for convection volumes with aspect ratios $\Gamma=L/H \le 4$ where $L$ is a 
horizontal extension (side length or diameter). Characteristic turnover times and the geometry of Lagrangian trajectories were analyzed 
numerically in closed non-rotating \cite{Emran2010} and in rotating cells \cite{Rajaei2016,Alards2017}. Laboratory experiments of 
turbulent convection pioneered the use of smart particles to measure local heat fluxes \cite{Gasteuil2007} along individual particle tracks. 
Complex three-dimensional particle tracking was applied to monitor Lagrangian acceleration statistics \cite{Ni2012} and the inhomogeneous flow 
behavior \cite{Liot2016}, respectively. The aspect ratios of the convection cells remained below $\Gamma\lesssim 6$ in all these 
cases. The present study is related to a Lagrangian analysis in a convection flow in a cell with larger aspect ratio of $\Gamma=16$ 
for which a large-scale organisation of the turbulent flow is expected. 

Recent three-dimensional numerical simulation studies in large-aspect-ratio layers in the Eulerian frame of reference showed that thermal 
plumes in turbulent convection form a web of connected ridge-like structures of ascending hot (descending cold) fluid from the bottom (top) 
plate \cite{Hartlep2005}. The corresponding large-scale fluid motion proceeds in form of circulation rolls \cite{Bailon2010,Emran2015} and 
the thermal plumes tend to cluster in the course of the dynamical evolution \cite{Parodi2004,Hardenberg2008}. The characteristic diameter of the 
large-scale circulation rolls, $\lambda_U/2$, or equivalently the characteristic distance between the strongest adjacent ridges of ascending 
and descending plumes, $\lambda_{\Theta}/2$, is a function of Rayleigh and Prandtl number \cite{Hartlep2003,Stevens2018,Pandey2018}. 
The resulting patterns, which are clearest revealed when a windowed time-averaging is performed in addition, are termed {\em turbulent superstructures} 
of convection. These superstructures evolve gradually for times larger than a characteristic time $\tau$ which is given by
$\tau\sim \ell/u_{\rm rms}$ with $\ell$ being the mean circumference of the large-scale rolls and $u_{\rm rms}$ the root mean square velocity 
\cite{Pandey2018}. Physically, this time $\tau$ is of the order of a typical turnover time of a fluid parcel in a large-scale circulation roll. 
A Lagrangian analysis of the plume and roll patterns in convective turbulence requires a different approach compared to previous studies 
in convective turbulence, namely a characterization of the coherent behavior of Lagrangian particles. 

Over the last two decades, a number of different concepts have been proposed that describe the notion of Lagrangian coherent behavior. For discussions and comparisons of the major current approaches we refer to refs. \cite{Allshouse_Peacock_2015,Hadjighasem2017}. The established approaches can be roughly divided into two different families. (1)
{\em Probabilistic methods} based on transfer operators are tailored to identify finite-time coherent sets as regions that are minimally 
dispersive over a finite-time span while moving with the flow.  For a regularization of the underlying optimization problem diffusion 
is artificially introduced, so that both advective and diffusive transport are minimized.  A uniform description of the different concepts 
in this framework can be found in \cite{FPG14}. (2) The dynamic objects central to {\em geometric methods} are Lagrangian coherent 
structures (LCS). They are defined as codimension-1 material surfaces that extremise a certain stretching or shearing quantity. Several heuristic diagnostic methods for the identification of LCS such as finite-time Lyapunov exponents (FTLE) 
but also mathematically sound variational criteria are reviewed in \cite{Haller_Rev_2015}. Whereas LCS form barriers to advective particle transport, the identification of optimal barriers to both advective and diffusive transport has been addressed only very recently in ref.~\cite{Haller2018}. 
Finally, another recent geometric characterization \cite{Froyland_2015} defines finite-time coherent sets as sets with minimal boundary to volume 
ratios under the action of the dynamics. This concept turns out to arise as the advective limit of the probabilistic framework and thus 
provides an analytical link between the two families of methods.

These established approaches are however less suited for our current study. Here, the aim is to study the emergence and structure of convection rolls, 
which are macroscopic entities. For this, particles are initialized in a two-dimensional plane near the bottom plate in a three-dimensional convective flow.
Whereas the geometric approach will identify the detailed turbulent skeleton including the small-scale structures such 
as vortices (see an FTLE-analysis in Sect.\ III.C;  we also refer to \cite{Green2007} for a study of a three-dimensional turbulent shear flow), the probabilistic framework -- although tailored for targeting large-scale 
structures in general -- is not immediately applicable either due to the two-dimensional seed of particles. Thus, an approach is required that can identify large-scale coherent behavior directly from \neu{the given} Lagrangian trajectory data. Recent promising works make use of spatio-temporal clustering algorithms applied to Lagrangian trajectory data
\cite{Froyland_Padberg_2015,Hadjighasem2016,Banisch2017,Schlueter2017,Schneide2017}, where the aim is to identify coherent sets as groups of trajectories that remain close and/or behave similarly in the time span under investigation.

Here, our focus will be on the latter approaches. In particular, we apply spectral graph theory \cite{Chung1997,Newman2010} and, for the first time in this context, the algorithm known as density-based spatial clustering applications with noise (DBSCAN) \cite{Ester1996} 
to analyze the time evolution of a Lagrangian particle ensemble
as a whole. To this end, the particle trajectories are composed to a network and its connection to the large-scale organization of a three-dimensional 
turbulent convection flow at large aspect ratio is studied. The set of $N_p$ individually advected Lagrangian particles at a time $t$ forms a set of 
vertices $\{v_1, \dots, v_{N_p}\} \in V$ of a weighted and undirected graph $G=(V,E,w)$. The vertices are connected by edges $\{e_1, \dots, e_M\}\in E$. 
The dynamical history of the particle ensemble up to time $t$ is encoded in weights $w$ which are assigned to each edge. Our applied trajectory
analysis follows in their steps the one of Hadjighasem {\it et al.} \cite{Hadjighasem2016} for Lagrangian vortex detection. 
The weights are calculated as the inverse of a time-averaged 
distance of mutual tracer trajectories and set to zero if this distance exceeds a threshold value. The latter is the sparsification step. By solving a balanced 
cut problem via an equivalent generalized eigenvalue problem of the Laplacian matrix of the graph \cite{Shi2000}, the network is decomposed 
into $k$ subgraphs or clusters. We relate the obtained clusters of the graph 
to superstructure rolls of the turbulent convection flow. The dispersion of the Lagrangian particles that start their evolution in close distance to each other
is an immanent property of a turbulent flow and can be quantified by the largest Lyapunov exponent. The standard spectral clustering analysis is 
applicable for the shorter-time evolution only, i.e., for time intervals $t\lesssim \tau$. We will compare the results to the corresponding finite-time Lyapunov exponent field, which measures local separation over a finite-time horizon. For times $t\gtrsim \tau$, a density-based clustering \cite{Ester1996} 
of pseudo-trajectories, which are obtained as time averages of Lagrangian trajectories, will be applied to probe the long-living turbulent superstructures. 
The latter analysis step gives us the most coherent subset of trajectories which are trapped for a long time in the core of the circulation rolls.   

Spatio-temporal clustering of Lagrangian particle trajectories has been applied recently for several flows with dominating vortex
structures, such as a Bickley jet on a plane \cite{Hadjighasem2016,Schlueter2017,Banisch2017}, two-dimensional polar vortex flow \cite{Schneide2017},
point vortex flows \cite{Nair2015}, and two-dimensional box turbulence with an inverse energy cascade \cite{Taira2016,Hadjighasem2017}. 
The present work comprises an application of such an analysis to a complex three-dimensional turbulent convection flow in an extended domain and 
the first use of DBSCAN in this context. Our main objective in this work is rather to apply these concepts for a Lagrangian analysis of turbulent 
superstructures in one example than to present comprehensive parameter studies at higher Rayleigh numbers or even larger aspect ratios, which 
are left for the future.

In the following Sec. II, we will summarize the numerical method and the basics of the applied graph theory. In Section III we will determine the 
characteristic scales of the superstructures in the Lagrangian frame of reference, report our results on the short-term analysis for times $t\lesssim \tau$ 
and the long-term analysis for $t\gtrsim \tau$. We conclude the work with a summarizing discussion in Sec. IV and give a brief outlook.

\section{Methods}
\subsection{Boussinesq approximation and numerical model}
We solve the three-dimensional equations of motion of turbulent convection in the Boussinesq approximation for which the fluid mass density is 
a linear function of the temperature. The equations are made dimensionless by using
height of the cell $H$, the free-fall velocity $U_f=\sqrt{g \alpha \Delta T H}$ and the imposed temperature difference $\Delta T$. 
Times are expressed in units of the free-fall time $T_f=H/U_f$.
The equations are given by
\begin{align}
\label{ceq}
{\bm \nabla}\cdot {\bm u}&=0\,,\\
\label{nseq}
\frac{\partial{\bm u}}{\partial  t}+({\bm u}\cdot{\bm\nabla}){\bm u}
&=-{\bm \nabla}  p+\sqrt{\frac{\mbox{Pr}}{\mbox{Ra}}} {\bm \nabla}^2{\bm u}+  T {\bm e}_z\,,\\
\frac{\partial  T}{\partial  t}+( {\bm u}\cdot {\bm \nabla})  T
&=\frac{1}{\sqrt{\mbox{RaPr}}} {\bm \nabla}^2  T\,,
\label{pseq}
\end{align}
with Rayleigh and Prandtl numbers that are given by Ra $=g\alpha\Delta T H^3/(\nu\kappa)$ and Pr $=\nu/\kappa$.
No-slip boundary conditions for the fluid (${\bf u}=0$)  are applied at all walls. The side walls are thermally insulated ($\partial T/\partial {\bf n}=0$) and 
the top and bottom plates are held at constant dimensionless temperatures $T=0$ and 1, respectively. These specific side wall boundary conditions are
taken in order to be able to compare the results to laboratory experiments which are currently in progress. In the following, we consider the case of 
Ra $=10^5$, Pr $=0.7$ and $\Gamma=16$. The simulation data are generated by the spectral element method nek5000 \cite{Fischer1997,Scheel2013}
with 441,000 spectral elements and a polynomial expansion up to 5th order which is parallelized with respect to all three space dimensions. The 
mean turbulent heat transfer is measured by the Nusselt number which is given by Nu $= 1 + \sqrt{\mbox{RaPr}} \langle u_z T\rangle_{V,t} = 4.1$, 
where the average of the heat flux is taken with respect to the whole domain $V$ and total time $t$. 

Lagrangian tracer particles are seeded into the simulation domain on a regular mesh with $N_p=512^2$ points at a height of $z=0.03$ above 
the bottom plate which is well inside the thermal boundary layer that has a mean thickness $\delta_T=1/(2\mbox{Nu})=0.121$. Each individual tracer particle is 
advected corresponding to 
\begin{equation}
\frac{d {\bm X}_i}{dt}={\bm u} ({\bm X}_i,t)\,,  
\label{lag}
\end{equation}
with $i=1\dots N_p$. The interpolation of the velocity field to the tracer position can be done spectrally in the present simulation code and is thus 
as accurate as the spectral expansion of the Eulerian fields itself. Time advection of tracers is done by a two-step Adams-Bashforth method. 
\begin{figure*}
\begin{center}
\includegraphics[width=1\textwidth]{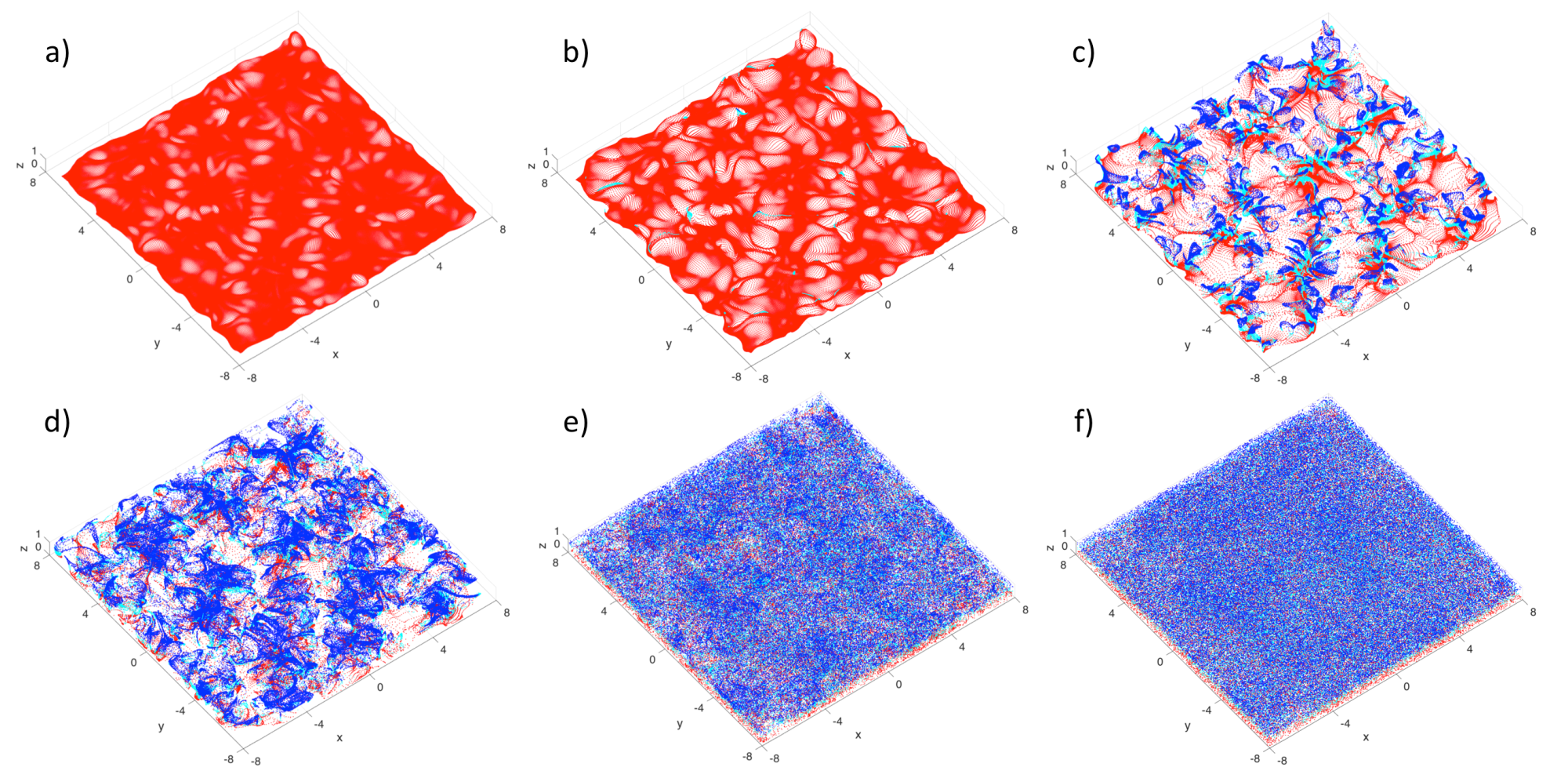}
\end{center}
\caption{Lagrangian tracer particle distributions in the turbulent convection cell at different times. Tracers with a vertical position of $0\le z< H/3$
are colored in red, $H/3\le z < 2H/3$ in cyan, and $2H/3\le z\le H$ in blue. Times in free-fall time units are $t=1.3\, T_f$ in (a), 2.6 $T_f$ in (b), 6.5 $T_f$ in (c),
10.4 $T_f$ in (d), 30.1 $T_f$ in (e), and 49.8 $T_f$ in (f).}
\label{cloud}
\end{figure*}
Figure \ref{cloud} displays the tracer distribution at different instants. Initially, the Lagrangian particles accumulate in regions of rising thermal 
plumes (see panels (a--c)) and are advected  to the top plate where they separate hyperbolically (see panel (d)) and move downward again, either in the same circulation roll or in different rolls. With progressing time tracer particles fill the whole convection layer as visible in panels (e,f). In the perspective view they appear as a well-mixed cloud that contains no more information about the large-scale patterns of the advecting flow. Our following analysis will show that information on the advecting flow structures is still obtainable from the tracer ensemble.

\subsection{Spectral analysis of Lagrangian tracer network}
In correspondence with \cite{Hadjighasem2016} the trajectory cluster analysis comprises the following steps: (i) graph construction from the trajectory points, (ii) determination of the weights at the edges, (iii) sparsification of the graph, (iv) solution of the generalized eigenvalue problem of
the graph Laplacian matrix and determination of the gap in the ordered eigenvalue spectrum, and (v) cluster detection by $k$-means clustering where $k$ 
indicates the first pronounced gap in the spectrum. 
The $N_p$ Lagrangian particle trajectories ${\bm X}_{i}(t)$ are sampled at discrete time instances $t_k=\{0, 1,\ldots, N_t\} \Delta t$. This 
information enters the construction of an undirected, weighted network with $n=N_p$ nodes or vertices. The weight $w_{ij}$ of the link between nodes 
$i$ and $j$ depends on the time-averaged distance $r_{ij}$ between the nodes \cite{Hadjighasem2016} which is given by 
\begin{equation}
r_{ij}=\frac{1}{N_t\Delta t}\int_{0}^{N_t\Delta t} |{\bm X}_i(t) - {\bm X}_j(t)| \,dt =\frac{1}{w_{ij}}\,.
\end{equation}
The weights thus contain the mutual trajectory histories up to $t=N_t\Delta t$ as an accumulated measure. A network with $n$ vertices has 
$M= n(n-1)/2$ edges if every node or vertex is connected with every other node. Self-connections are excluded.
In order to sparsify the network (and thus the related adjacency or similarity matrix), we only consider links with weight $w_{ij} > 1/\epsilon$ 
with $\epsilon>0$. This network can be uniquely described by the similarity matrix $W \in \mathbb{R}^{n,n}$ with
\begin{equation}
W_{ij}=
\begin{cases} w_{ij},& i\neq j,\; r_{ij} < \epsilon \\ 
0, & \mbox{else} \end{cases}. \label{eq:W}
\end{equation}
The choice of $\epsilon$ is discussed later in the text in Section III. The degree matrix $D$, given by $D_{ii}=\sum_{j=1}^n W_{ij}$, is 
a diagonal matrix with node degrees $d_i=D_{ii}$ as entries. It can be used to calculate the non-normalized graph 
Laplacian $L = D-W$. As shown by Shi and Malik \cite{Shi2000}, the solution of the generalized 
eigenvalue problem 
\begin{figure*}
\begin{center}
\includegraphics[scale=0.7]{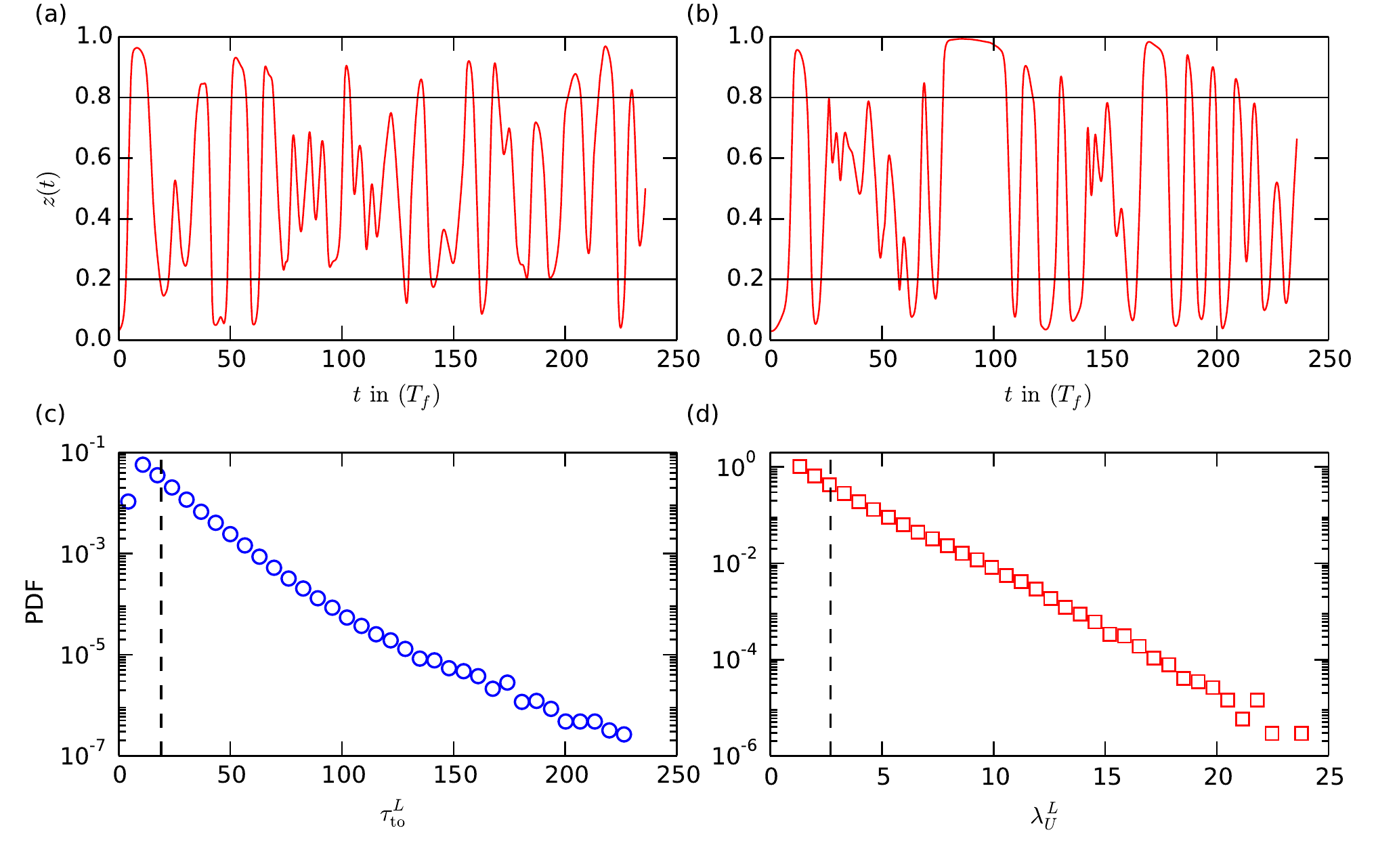}
\end{center}
\caption{Characteristic Lagrangian times and scales of turbulent superstructures. (a,b) Typical Lagrangian particle tracks showing the vertical particle
coordinate $z(t)$ for individual tracers no. 2 in (a) and no. 2045 in (b). The horizontal lines at $z=0.2$ and $0.8 H$ in both panels indicate the heights
that were used to determine the Lagrangian turnover times.  (c) Probability density function (PDF) of Lagrangian turnover time $\tau^L_{\rm to}$.
The dashed vertical line indicates the mean Lagrangian turnover time  
$\overline{\tau}^L_{\rm to}= (19.1 \pm 12.8) T_f$. The root mean square (rms) value from this statistics is $23.0 T_f$. (d) PDF of the characteristic 
Lagrangian scale $\lambda_U^L$. The vertical line indicates again the mean with  $\overline{\lambda}^L_{U}= (2.7 \pm 0.9) H$. This stands for the diameter of 
a pair of rolls. The rms value is $3.2 H$.}
\label{tracks}
\end{figure*}
\begin{equation}
Lv=\lambda Dv, \label{eq:ev_problem}
\end{equation}
serves as indicator for the subdivision of the graph into clusters. The eigenvalues are non-negative and real, $0=\lambda_1 \leq 
\lambda_2 \leq \ldots \leq \lambda_{n}$, with the corresponding eigenvectors $v_1, \ldots, v_n$. In particular, the eigenvector 
corresponding to the second smallest eigenvalue of (\ref{eq:ev_problem}) approximates the solution of the normalized cut (Ncut) 
problem introduced by \cite{Shi2000} which seeks to minimize the connectivity between clusters while maximizing simultaneously the 
connectivity within clusters. Eigenvectors corresponding to the next smallest eigenvalues can be used to further subdivide the graph. 
The number of leading eigenvalues close to zero determines the number of nearly decoupled communities in the network 
\cite{Fiedler1973,vonLuxburg2007}. Such nearly decoupled subgraphs correspond to bundles of trajectories that are internally well 
connected but only loosely tied to other 
trajectories, an indication of coherent behavior (see also \cite{Hadjighasem2016}). We apply a $k$-means clustering algorithm on the 
eigenvectors corresponding to the $k$ leading eigenvalues in order to extract $k$ clusters \cite{Goodfellow2016}. 
The number of leading eigenvalues is determined by a spectral gap heuristics.

\section{Results}
\subsection{Characteristic time and length scales of large-scale patterns in Lagrangian frame}
The first question to answer is the one on the agreement of the characteristic pattern scales in the Eulerian and Lagrangian frames of reference. Both,
the characteristic time and length scales are of central interest for a separation of the fast, small-scale from the gradual, large-scale dynamics.  
In correspondence with \cite{Pandey2018}, we determined the characteristic time of the turbulent superstructures in the Eulerian frame of 
reference with $\tau\approx 3\ell/u_{\rm rms} = 59 T_f$ for the present data set. The characteristic Eulerian length scale of the temperature patterns 
was determined to $\lambda_{\Theta}=4$, either by a Fourier analysis or the evaluation of the spatial correlations. The scale of a pair of counterrotating circulation rolls, $\lambda_U$, was of comparable size. 
\begin{figure}
\begin{center}
\includegraphics[scale=0.5]{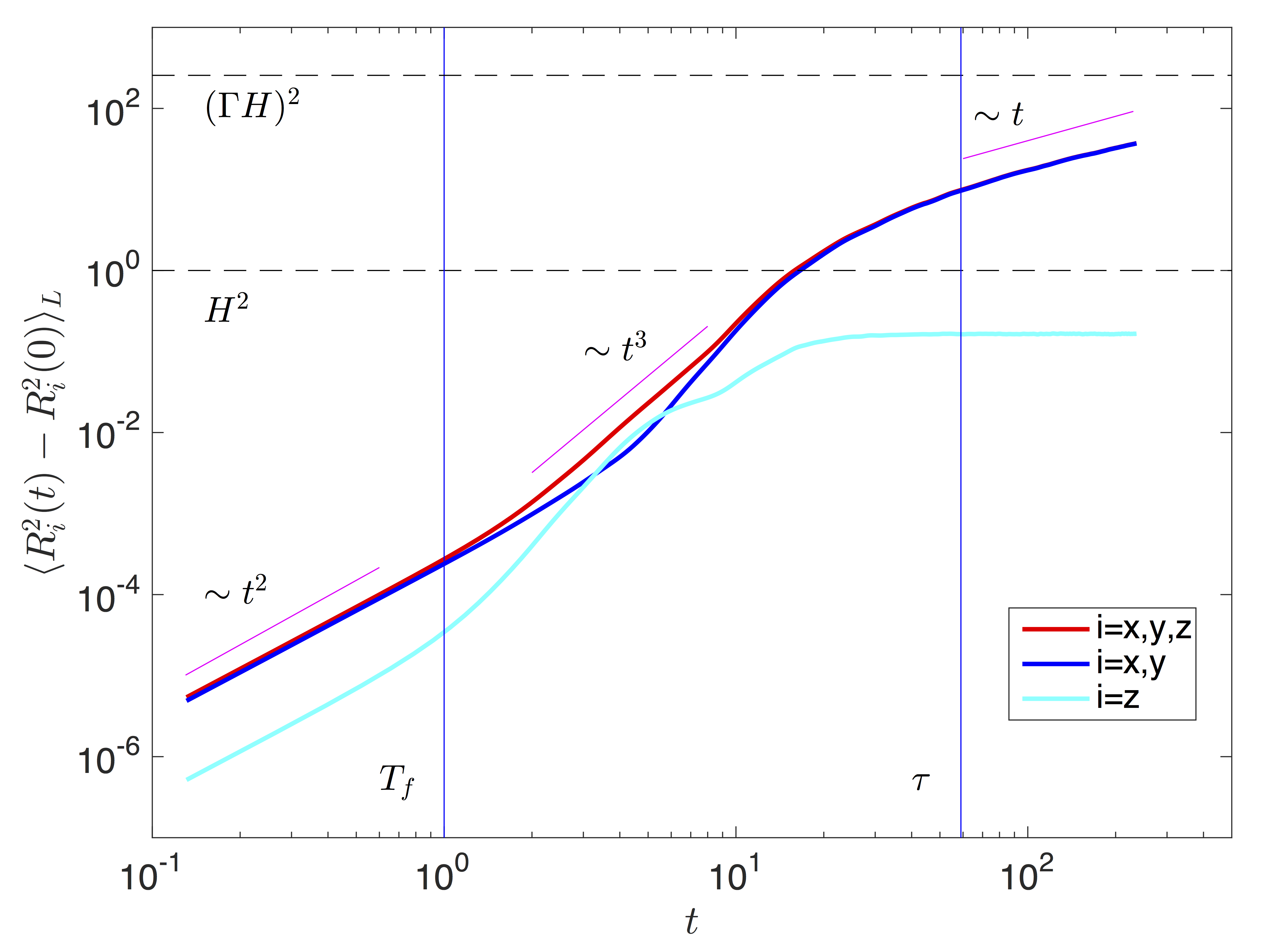}
\end{center}
\caption{Lagrangian tracer pair dispersion versus time. Ballistic, Richardson-like intermediate and diffusive scaling are indicated in addition to
all relevant times and outer scales in the problem. The total dispersion is shown together with its two contributions, the vertical and horizontal pair 
dispersions. Times are given in units of the free-fall time $T_f=H/U_f$, pair dispersion in units of $H^2$. Again $\tau\approx\tau^L$.}
\label{pair}
\end{figure}

Here, the large-scale structure can be probed directly on the basis of the Lagrangian particle tracks. The analysis is summarized in Fig.~\ref{tracks}.
Panels (a) and (b) display the vertical coordinate of two typical Lagrangian particle tracks. As visible, the tracers do not follow a perfect 
upwelling-downwelling motion, remain partly trapped close to the top or bottom plates as well as in the center of the convection layer. We measure 
the turnover time for each individual Lagrangian particle as the time it takes to rise from heights $z<0.2$ to $z>0.8$ and back down again. These heights
are slightly above and below the lower and upper thermal boundary layers, respectively. Recall again that the thermal boundary layer thickness is 
$\delta_T=0.121$. The corresponding probability density function (PDF) of the Lagrangian turnover time $\tau^L_{\rm to}$ is given in Fig.~\ref{tracks}(c).
The PDF is characterized by a fat (stretched exponential) tail that implies the high probability of long excursions of the Lagrangian tracer particle in the bulk. The mean value is found to be
$\overline{\tau}^L_{\rm to}= (19.1 \pm 12.8) T_f$ and thus agrees very well with the Eulerian analysis, $\tau/3\approx 19.7\,T_f$. 
It can now also be seen that the prefactor of 3, which we have chosen in the Eulerian analysis in \cite{Pandey2018}, is justified by the 
tail of the distribution. Turnovers 
can take as long as 200 $T_f$ and even more. In ref. \cite{Pandey2018}, it was shown that somewhat smaller or larger prefactors than 3 lead to very similar 
results. We conclude that 
\begin{equation}
\overline{\tau}^L_{\rm to}\approx \frac{1}{3} \tau \quad\quad\mbox{and thus}\quad\quad \overline{\tau}^L\approx \tau
\label{scale1}
\end{equation}
follows consistently from our analysis.       

For the calculation of a characteristic spatial scale in the Lagrangian frame, we take the midplane of the convection layer at $z=0.5$ and 
determine the distance of the points of subsequent rising and falling events of each Lagrangian particle track. This distance corresponds to 
$\lambda_U^L/2$ since the wavelength is taken for a pair of circulation rolls. The corresponding distribution is plotted in Fig.~\ref{tracks}(d). 
Again, we detect an exponential tail that suggests a wide range of possible scales. The mean value is 
$\overline{\lambda}^L_{U}= (2.7 \pm 0.9) H$. This number is a bit smaller than the Eulerian value of 4 which can be attributed to trajectories that get trapped
in the core regions of the superstructure rolls for longer times. The large variability is seen from the error bar in the mean values. We conclude that 
\begin{equation}
\overline{\lambda}^L_{U}\lesssim \lambda_U
\label{scale2}
\end{equation}
follows from our analysis and that both scales of the superstructures consistently match with their Eulerian counterparts.       

In section II B we introduced already the parameter $\epsilon$ for the sparsification of the graph which is associated with the time-averaged 
distance between Lagrangian trajectories. We have chosen a value of $\epsilon=0.75$ throughout our analysis. In the absence of a rigorous criterion for the
choice of this parameter, we provide physical arguments in the following. The cut parameter $\epsilon$ should be bounded from below by 
the thermal boundary layer thickness $\delta_T$ which is comparable to the width of the stems of detaching line-like thermal plumes
\cite{Theerthan1998}. In a rising (or falling) 
thermal plume two Lagrangian particles will display a correlated upward (or downward) motion. The thickness of the thermal boundary 
layer varies locally and forms a distribution with $\delta_T$ as the mean. As shown in ref. \cite{Scheel2014}, such a local boundary layer thickness 
distribution has typically an extended tail to larger scales.
Furthermore the thermal plumes disperse as they move to the bulk. This causes an additional increase of their thickness. On the large-scale end,
the cut parameter $\epsilon$ should be bounded from above by $\lambda_{\Theta}/2\approx \lambda_U/2$, which is about 2 in our case at hand.

\subsection{Connection to Lagrangian particle pair dispersion}
It is interesting to relate the detected time scale $\tau\approx \tau^L$ of the superstructures to the Lagrangian particle pair dispersion. All particle pairs 
had the same initial separation of $R_0=0.005$. The pair dispersion is defined as the mean square of the particle pair distance, the latter 
of which is given by ${\bm R}(t)={\bm X}_1(t)-{\bm X}_2(t)$.  It can be decomposed into a horizontal and vertical 
contribution \cite{Schumacher2009}
\begin{equation}
{\bm R}^2(t) = {\bm R}^2_{xy}(t) + R^2_z(t){\bm e}_z\,.
\label{disp}
\end{equation}
In  Fig.~\ref{pair} we display the pair dispersion $\langle {\bm R}^2(t)-{\bm R}_0^2\rangle_L$ versus time with ${\bm R}_0^2={\bm R}^2(0)$. 
The symbol $\langle\cdot\rangle_L$
stands for an average over all $N_p$ trajectories. Times $t\lesssim T_f$ are characterized by the ballistic separation of tracers within a pair. This 
follows by a Taylor expansion and results in a scaling  of $\langle {\bm R}^2(t)-{\bm R}_0^2\rangle_L\sim t^2$. For $t\gtrsim 10 T_f$ the vertical pair dispersion 
$\langle R_z^2(t)-R_{z,0}^2\rangle_L$ saturates to a constant that is smaller than $H^2$. In a crossover period between $t\sim T_f$ and 10 $T_f$, the total pair 
dispersion follows a Richardson-type scaling that grows with $t^3$. It is the horizontal dispersion $\langle {\bm R}^2_{xy}(t)\rangle_L$ that 
eventually takes over and dominates the long-time behavior. The long-time limit of the pair dispersion is the diffusion limit for which 
$\langle {\bm R}^2_{xy}(t)\rangle_L \sim t$ \cite{Sawford2008}. Figure~\ref{pair} shows clearly that the beginning of this regime coincides with the 
characteristic time scale of the turbulent superstructures,
$\tau$. From the perspective of an individual Lagrangian tracer particle, the motion within a particle pair is then almost decorrelated and 
requires an additional clustering. 
In the following, we conduct the network analysis in respect of the determined time scale and divide it into a short-term ($t\lesssim\tau$) and long-term
($t\gtrsim\tau$) trajectory clustering.

\subsection{Short-term evolution of the Lagrangian network}
\begin{figure*}
\begin{center}
\includegraphics[width=0.95\textwidth]{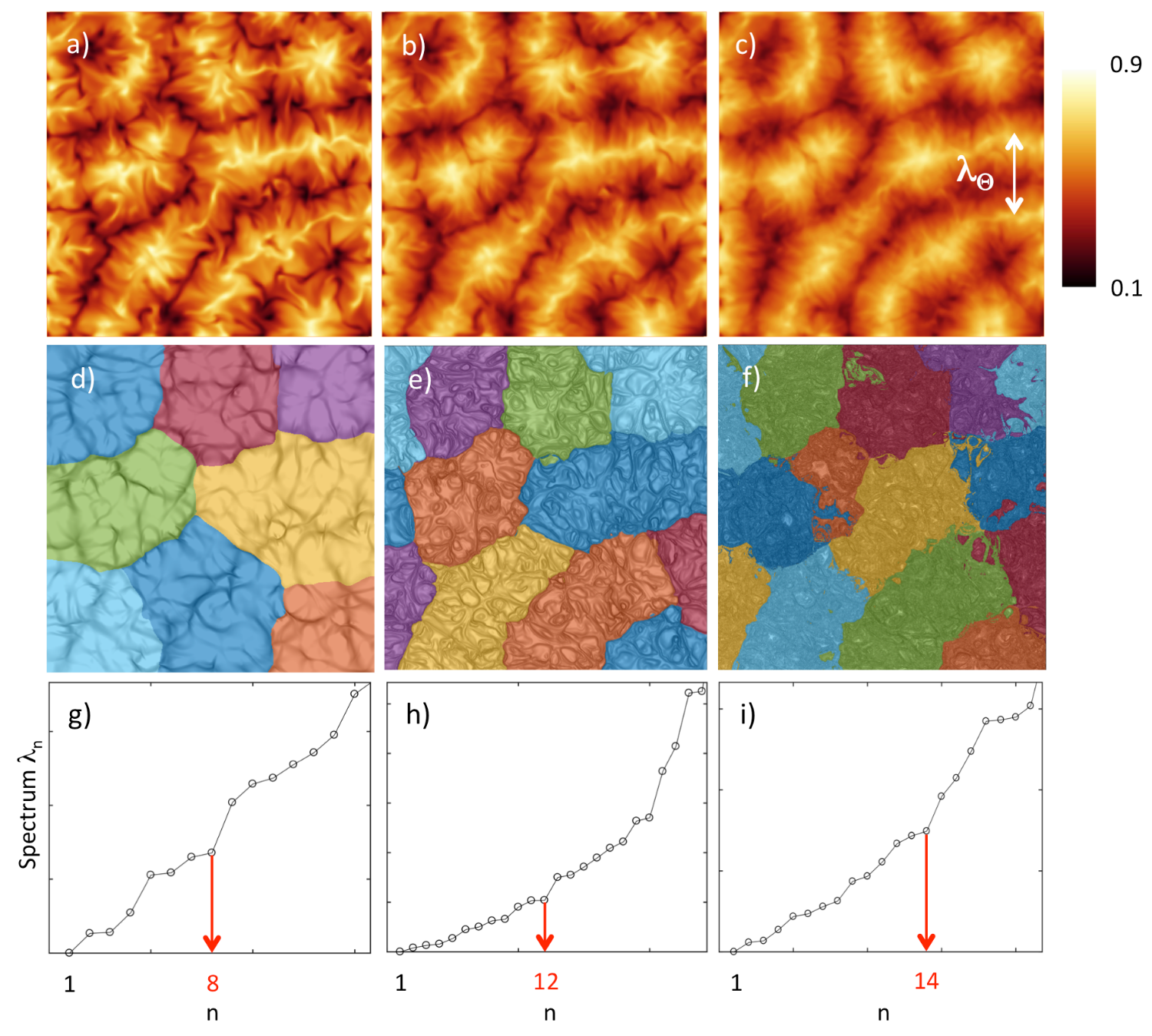}
\end{center}
\caption{Time evolution of the Lagrangian superstructures. (a--c) Time-averaged temperature field in the midplane. Averaging times are 
$N_t\Delta t$ = 2.6, 10.4, and 30.1 free-fall time units, respectively. In panel (c), the characteristic scale $\lambda_{\Theta}$ of the turbulent 
superstructures is indicated by a double arrow. (d--f)  Lagrangian trajectory clusters obtained from the leading eigenvectors of the graph Laplacian. Particles 
that belong to the same spectral cluster at time $N_t \Delta t$ are colored equally. The background contours are the ridges of the maximum finite-time
Lyapunov exponent \eqref{eq:ftle}. Ridges and clusters are indicated with respect to the initial Lagrangian particle position. (g--i) Corresponding eigenvalue spectra of the graph  Laplacian. 
The spectral gap between eigenvalues no. 8-9, 12-13, and 14-15 is used to detect $k$ = 8, 12, and 14 trajectory clusters 
by the $k$-means clustering algorithm, respectively. The cut-off parameter is $\epsilon=0.75$. }
\label{graph}
\end{figure*}

Figure \ref{graph} summarizes our spectral graph analysis for times $t\lesssim \tau^L$.  All particles start inside the thermal boundary layer 
at a distance $z\approx \delta_T/4$ from the bottom wall. Panels (a)--(c) show the temperature at the midplane $z_0=1/2$ which is successively 
longer time-averaged for three different time intervals $N_t\Delta t$ in correspondence with
\begin{equation}
\Theta(x,y,z_0; N_t\Delta t) =\frac{1}{N_t\Delta t} \int_{0}^{N_t\Delta t} T(x,y,z_0,t^{\prime}) dt^{\prime}\,. 
\label{timeaverage}
\end{equation}
This definition of the time-averaged temperature is similar to the one used in ref.~\cite{Pandey2018}. 
The characteristic scale of the turbulent superstructures of the temperature field $T$ which is denoted as $\lambda_{\Theta}$
is indicated in Fig.~\ref{graph}(c). 

The two lower rows of Fig. \ref{graph} display the results of the spectral analysis and the subsequent clustering for these times.
The cut-off parameter $\epsilon=0.75$.  
Panels (d)--(f) depict the different clusters. We plot the initial position of the Lagrangian particles that belong to the same cluster in the 
same color. The background texture stands for the field of the maximum FTLE advanced over the same time 
interval \cite{Haller_Rev_2015}.  The FTLE for some initial particle position ${\bm X}_0={\bm X}(0)$ over the time span $[0, N_t\Delta t]$ (i.e. $t = N_t\Delta t$) is 
defined as
\begin{equation} \label{eq:ftle}
\sigma_{t}({\bm X}_0)=\frac{1}{2N_t \Delta t}\ln \lambda_{\rm max} \left(\Phi_t({\bm X}_0)\right),
\end{equation}
where 
\begin{equation}
\Phi_t({\bm X}_0):=\left(\frac{d}{d {\bm X}_0}{\bm X}(t)\right)^{T}\cdot\frac{d}{d {\bm X}_0}{\bm X}(t)\,.
\end{equation}
denotes the Cauchy-Green strain tensor and $\lambda_{\rm max}$ the largest eigenvalue of this tensor.

While the FTLE field initially picks up isolated structures that appear to be the signature of the boundaries between convection roles (cf. \cite{Tang2011}), it displays ever finer textures with increasing time while the cluster number remains almost the same. 
At early times, the FTLE ridges coincide partly with the cluster boundaries. The best agreement of the cluster boundaries and most prominent FTLE ridges 
is achieved for the time interval of $10.4$ free-fall time units, panel (e). Panels (g)--(i) of the same figure show the eigenvalue 
spectra of the graph Laplacian matrix $L$ at the corresponding time and highlight the spectral gap that suggests the segmentation into $k$ 
clusters by a $k$-means clustering algorithm \cite{Goodfellow2016}. The resulting trajectory clusters agree rather well with the appearing Eulerian 
superstructure patterns. This can be seen by a comparison of the first and second rows of Fig. \ref{graph}. However, the trajectory clusters 
become increasingly fragmented as the monitoring time $N_t\Delta t$ approaches $\tau^L$ (not shown). Turbulent dispersion separates an 
increasing fraction of pair trajectories as time progresses. This eventually also limits the trajectory-based clustering algorithm. 

\begin{figure*}
\begin{center}
\includegraphics[width=0.9\textwidth]{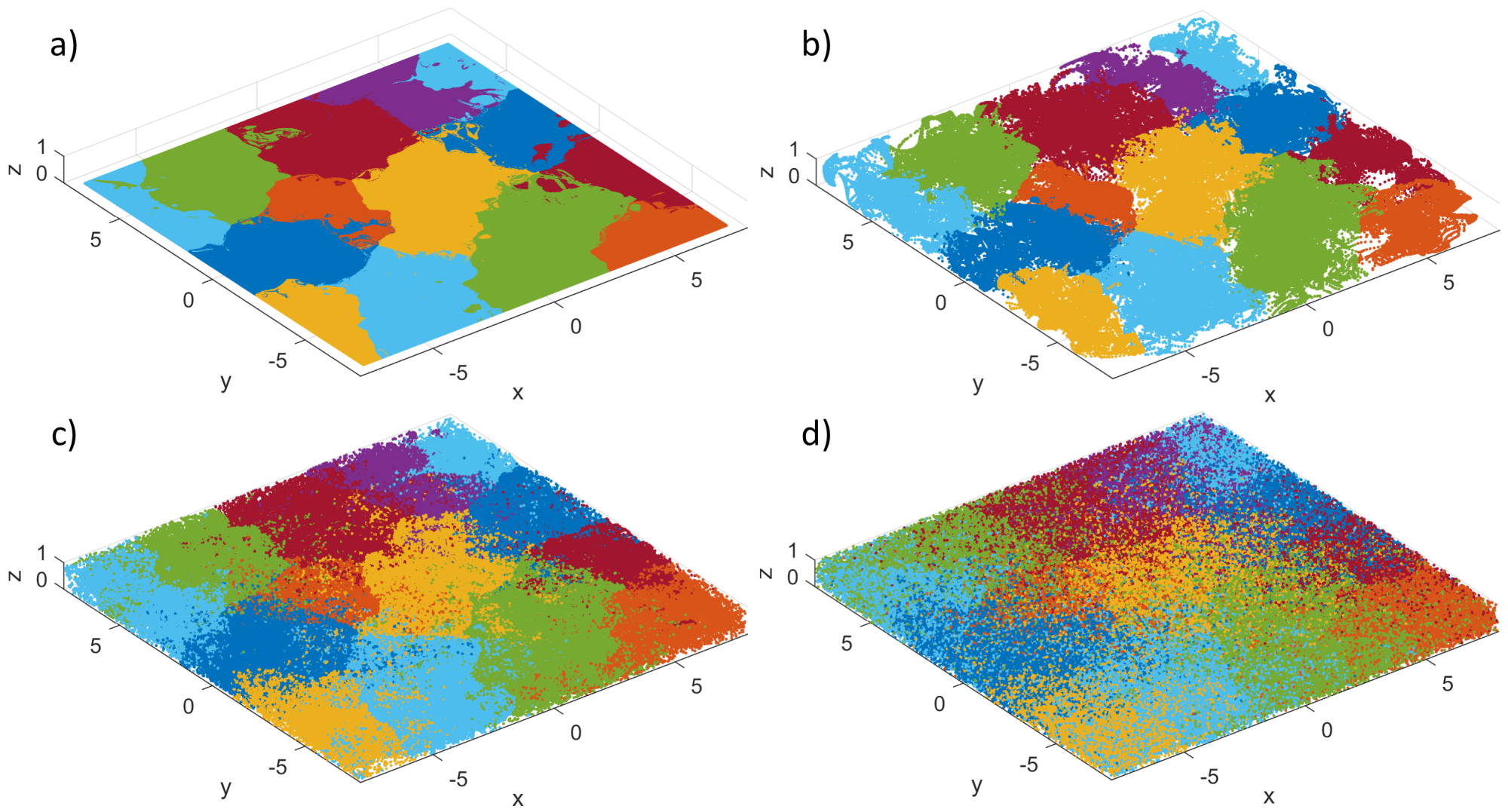}
\end{center}
\caption{Perspective view of the whole Lagrangian tracer ensemble at times $t=0\,T_f$ in (a), 10.4 $T_f$ in (b), 30.1 $T_f$ in (c), and 69.4 $T_f$
in panel (d). The particles are colored with respect to their trajectory cluster membership as computed for the case $N_t\Delta t= 30.1\, T_f$ (cf. Fig. \ref{graph} (f)).}
\label{cluster_evol}
\end{figure*}

The temporal evolution of the Lagrangian tracers within the clusters is displayed in Fig. \ref{cluster_evol}. We show perspective plots of the full
tracer ensemble at four different times, $N_t\Delta t = 0,\,10.4,\,30.1$ and 69.5 $T_f$. In all four panels of this figure we color each Lagrangian tracer 
with respect to the trajectory cluster that was identified for the averaging time $N_t\Delta t=30.1\,T_f$ (see also Fig. \ref{graph} (f)) for this particle. The figure 
demonstrates that the majority of the Lagrangian
tracers stay in the cluster up to times $t = 30.1\, T_f$. Panel \ref{cluster_evol}(d) shows in addition that even for more than twice this time 
(and $t\simeq \tau$) the coarse 
structure of the clusters is still identifiable although a significant amount of the Lagrangian particles has been transported to other phase space regions due to turbulent dispersion.
 
We carried out the same analysis using different $\epsilon$ and  different numbers of tracer particles, respectively, leading to similar 
results for the number and distribution of clusters (not shown).
 
\subsection{Long-term evolution by density-based clustering of pseudo-trajectories}
\begin{figure*}
\begin{center}
\includegraphics[scale=0.15]{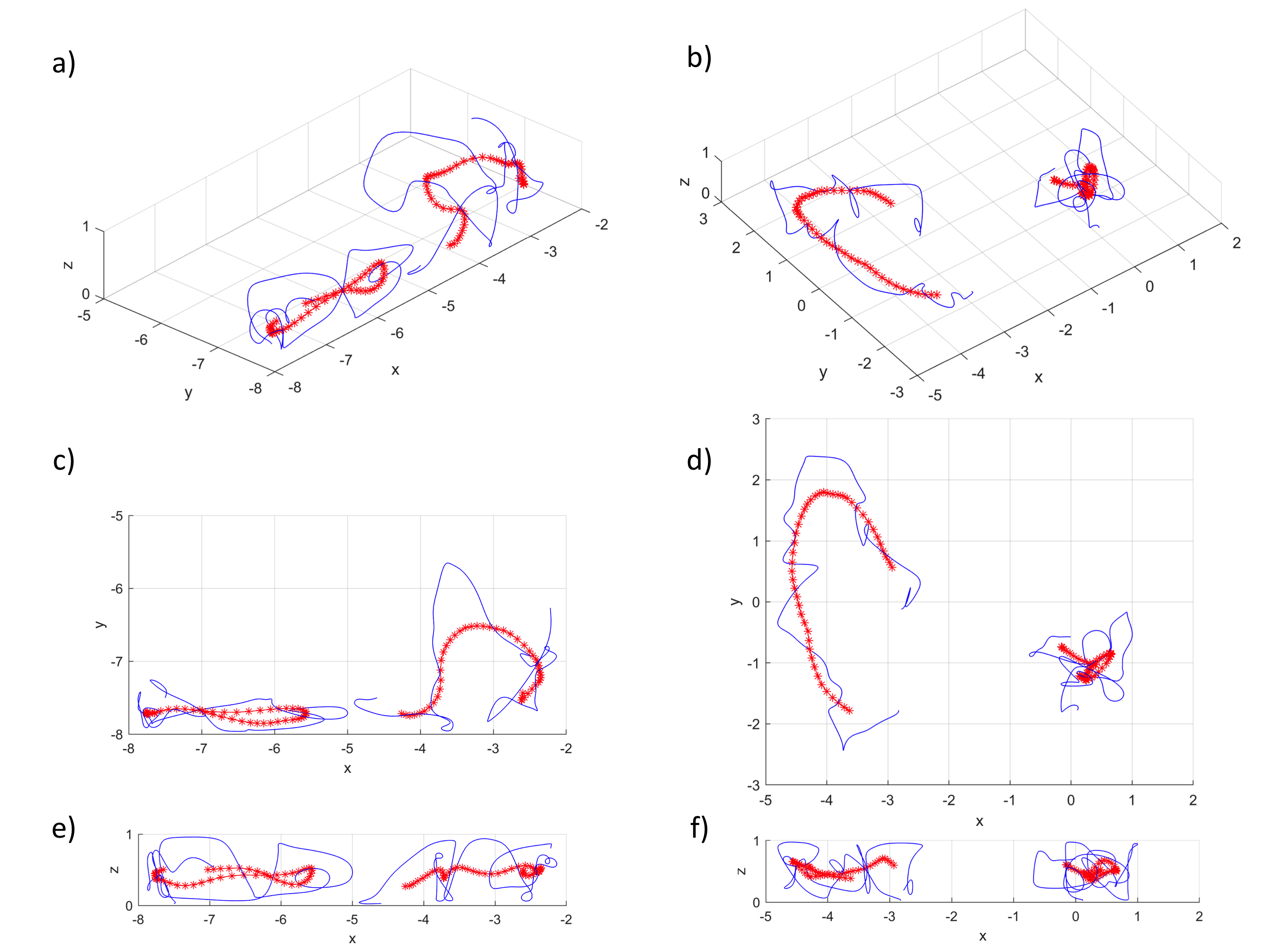}
\end{center}
\caption{Time evolution of four individual trajectories (in blue) and their corresponding pseudo-trajectories (in red) plotted for the 
time interval $t/T_f \in$ [0, 80]. The asterisks at the pseudo-trajectories indicate the positions $\overline{\bm X}(n \hat{\tau})$ for $n$ integer. Two trajectories 
are displayed in a three-dimensional perspective (panel (a)) and in two projections in panels (c,e). The same sequence follows for another set of two trajectories  
in panel (b) and panels (d,f), respectively.}
\label{pseudotrajectory}
\end{figure*}

The long-term evolution for times $t\gtrsim \tau^L$ requires the construction of pseudo-trajectories from the Lagrangian particle tracks which will
be described in the following. Lagrangian pseudo-trajectories are defined as 
\begin{equation}
\overline{\bm X}(n \hat{\tau} +\tau^L_{\rm to}) = \frac{1}{\overline{\tau}^L_{\rm to}} \int_{n \hat\tau}^{n \hat\tau +\overline{\tau}^L_{\rm to}}
\,{\bm X}(t^{\prime})\,dt^{\prime}\,.
\label{pseudo}
\end{equation}
with $n=0,1,\dots, n_{\rm end}$ and $\hat\tau = 1.31 T_f$ (which corresponds to 200 integration time steps). 
It turns out that we have to apply here again a coarse-graining procedure to reveal the superstructures. In a nutshell, the 
time average of the turbulence fields in the Eulerian case corresponds with the time average of the pseudo-trajectories in the Lagrangian 
frame of reference. The time scale $\overline{\tau}^L_{\rm to}$ is the mean turnover time of a Lagrangian particle in a superstructure roll as determined 
in Sec. III A. In free-fall time units, we therefore take $\tau^L_{\rm to}\approx \tau=19.7 T_f$ for the given set. The time-averaged position of a Lagrangian 
particle should thus coincide for 19 to 20 $T_f$ with the center of a superstructure roll. Many trajectories will not meet this stringent selection 
procedure due to the ongoing turbulent dispersion of particles that have been originally closely together.
Figure \ref{pseudotrajectory} displays the time evolution of four individual Lagrangian trajectories together with their corresponding pseudo-trajectories 
for a time interval $t/T_f \in$ [0, 80]. Clearly visible is that the pseudo-trajectories are found indeed in the bulk of the convection layer (see again panels 
\ref{pseudotrajectory} (e,f)).

Based on the obtained pseudo-trajectories, we apply the DBSCAN algorithm 
by Ester {\it et al.} \cite{Ester1996}. Considering a set of points $D$, DBSCAN evaluates the $\epsilon$--neighborhood of each point 
in the set $D$. A core point $P$ of a cluster ${\cal C}$ is a point whose $\epsilon$--neighborhood contains at least MinPts points, thus determining 
a density in the data space. All points in the $\epsilon$--neighborhood of $P$ are assigned to cluster ${\cal C}$. They can be core points such as 
point $P$ itself or border points of the cluster. The number of points in the $\epsilon$--neighborhood of border points of a cluster is smaller 
than the limit MinPts. Subsequently, this process is repeated for all core points of cluster ${\cal C}$. The set of points in the database which are not 
assigned to any cluster is defined as noise. We extend this algorithm to our spatio-temporal dataset by interpreting the pseudo-trajectories as points and using the 
dynamical distance $r_{ij}$ as distance function. 

\begin{figure*}
\begin{center}
\includegraphics[width=1\textwidth]{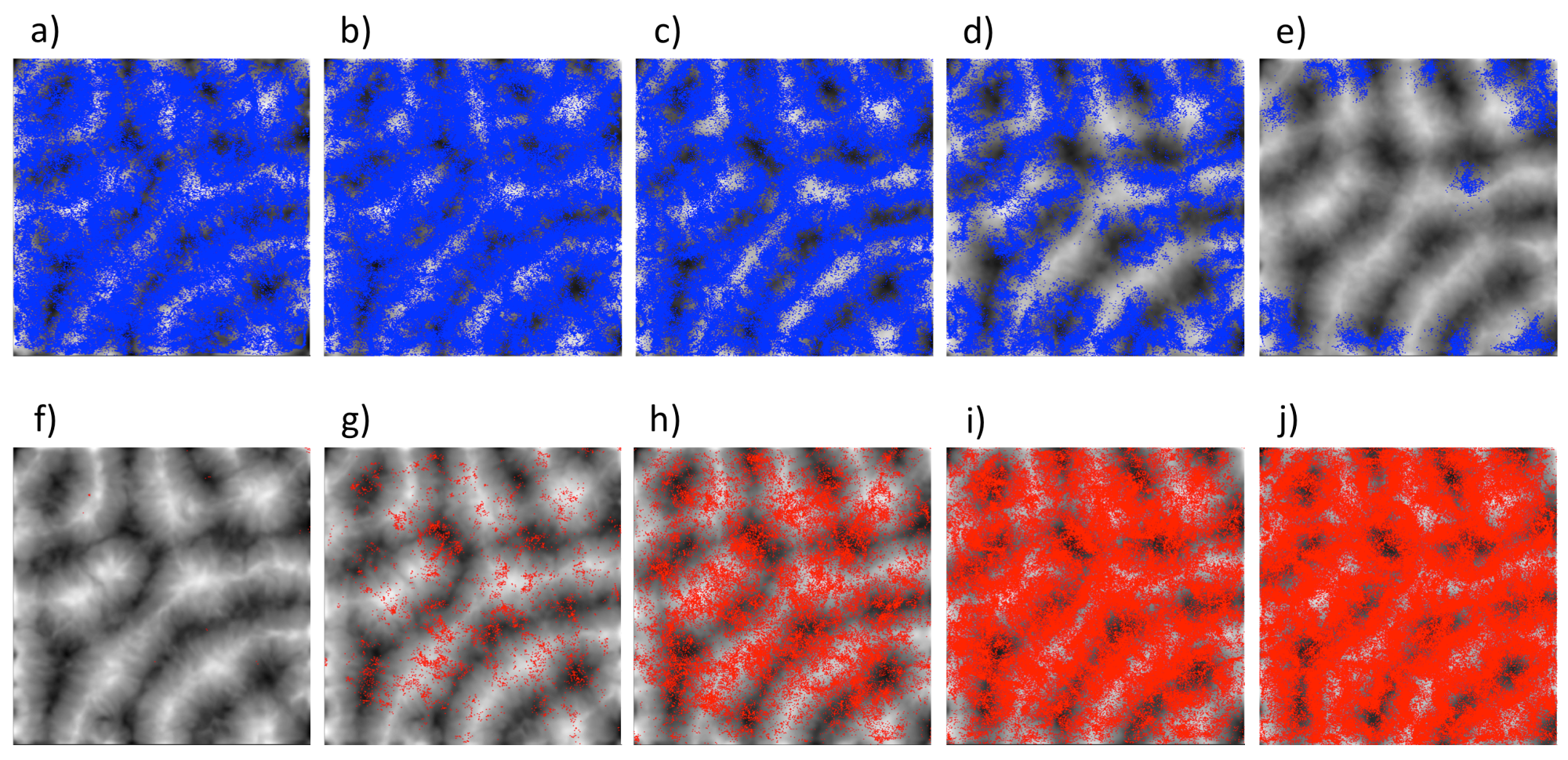}
\end{center}
\caption{Time evolution of the Lagrangian pseudo-trajectories for total integration times $T_{\rm end} = 40 T_f\approx 2\tau^L/3$ (a,f), 
$60 T_f \approx \tau^L$ (b,g), $80 T_f \approx 4\tau^L/3$ (c,h), $100 T_f \approx 5\tau^L/3$ (d,i), and $120 T_f\approx 2\tau^L$ 
(e,j). The panels in the upper row of the figure show the midpoints of the trajectories that belong to the clusters as given in Table~\ref{tab:dbscan}. The panels in the bottom row show noisy trajectories.}
\label{dbscan}
\end{figure*}

The resulting clustering into dense regions (clusters) and sparse regions (noise) for different integration times 
$T_{\rm end} = n_{\rm end} \hat{\tau} +\tau^L_{\rm to}$ using a value of MinPts = $500$ is 
summarized in Fig. \ref{dbscan} and Table \ref{tab:dbscan}. We will take again $\epsilon=0.75$ as in section III C for the short-term analysis.
The plots show the midpoints of the pseudo-trajectories, $n \approx n_{\rm end}/2$, according to the clustering in blue (dense regions) or red (sparse regions) over the time-averaged 
temperature field in midplane in gray. The number of noisy pseudo-trajectories increases with increasing time. These trajectories contribute to transport in the system as they mingle extensively with other trajectories during the considered time interval and move up- and downward.
 
The density arising from the chosen combination of the parameters $\epsilon$ and MinPts is too small to distinguish between mixing and 
non-mixing trajectories for the selected shortest time interval of $40 T_f$, as seen in the first row of Table \ref{tab:dbscan}. We have found 
that in this case an increase of MinPts to $1000$ leads to the detection of $\sim 2450$ noisy trajectories which is still less than 1 per cent of 
the total number of $N_p$ trajectories. Only for times $t > \tau^L$, the number of noisy trajectories grows.
Compared to trajectories in dense regions, which cluster between adjacent ridges of thermal plumes, noisy trajectories primarily assemble at 
ridge positions as seen in panels (b)--(d) and (g)--(i) of Fig. \ref{dbscan}. The noisy trajectories are thus found exactly where the strongest up- or 
downwelling motion and a subsequent trajectory separation is present. 

We conclude furthermore that the detected clusters are trajectories that remain trapped within the superstructure circulation rolls for a considerable 
part of the examined time interval. The DBSCAN algorithm detects thus exactly those pseudo-trajectories that contribute least to the transport.  
For even longer times, $t\gtrsim 2\tau$, the analysis in the present case is affected by the side wall boundary conditions and thus stopped. This 
is seen in panel (e) of Fig. \ref{dbscan} where the trajectories cluster at the side walls. We suspect that this effect will be absent when
periodic boundary conditions at the side walls are used. We leave this specific task for our future work.
\begin{table}
\begin{ruledtabular}
\begin{center}
\begin{tabular}{ccc}
 Time $T_{\rm end}$(in $T_f$)   &  No. of pseudo-trajectories in dense regions  &  No. of clusters \\
\hline
 40 & 262112 & 1 \\
 60 & 255047 & 1 \\
 80 & 210345 & 1 \\
 100 & 105551 & 29 \\
 120 & 17601 & 12 \\
\end{tabular}  
\caption{Results of the DBSCAN clustering for different time spans in units of the free fall time $T_f$. The threshold value that determines the $\epsilon$-neighborhood of a trajectory was chosen to $\epsilon=0.75$. The threshold to obtain a cluster was set in the analysis to MinPts = 500.}
\label{tab:dbscan}
\end{center}
\end{ruledtabular}
\end{table}

\section{Conclusions}
We conducted a Lagrangian analysis in a horizontally extended turbulent Rayleigh-B\'enard convection flow. The central motivation of our work was 
to probe the large-scale patterns in this complex three-dimensional flow, the turbulent superstructures, and compare the findings with an analysis 
in the complementary Eulerian frame of reference, as presented in ref. \cite{Pandey2018}. We confirm that the characteristic scales of the 
superstructures in the Eulerian and Lagrangian frames of reference agree. It is also shown that this characteristic time scale $\tau^L$ falls into the 
beginning of the Taylor regime of pair dispersion.   

The Lagrangian detection of turbulent superstructures calls for an approach that studies the Lagrangian particle ensemble as a whole. 
This was obtained here by application of network (or graph) theory-based algorithms that were suggested in \cite{Hadjighasem2016}
and determine the spectrum of a graph Laplacian matrix 
$L=D-W$ in 
which the information on the mean mutual distance is encoded in the edge weights. The spectral gap of $L$ indicates the number of clusters that can
be determined by a standard clustering algorithm subsequently. These resulting clusters are found to agree well with the superstructure patterns,
such as those of temperature shown in Fig. \ref{graph}. The method
works well up to time $\tau$, whereas, notably, the FTLE field only reveals isolated ridges of strong separation for times $\ll \tau$. For longer times the dominating fraction of Lagrangian particles are decorrelated (what defines the beginning of the 
Taylor dispersion regime). In this case a time averaging of the Lagrangian trajectories is required in order to distinguish mixing and non-mixing 
trajectories. We defined  therefore pseudo-trajectories. The density-based clustering algorithm DBSCAN is applied to find pseudo-trajectories that remain closely together for integration times up to $2\tau$. 
These pseudo-trajectory clusters appear to coincide with the center of circulation rolls in the Eulerian frame of reference. The clustering 
algorithm detects those trajectories which contribute least to the turbulent transport of heat from the bottom to the top thus forming a Lagrangian
coherent set.

As we stated in the introduction, the present work is understood as a first step to work out the Lagrangian concepts to describe large-scale patterns
in turbulent convection. Comprehensive parameter studies at different Prandtl numbers and for larger aspect ratios and Rayleigh numbers will be a part of the 
future work on this subject. This has to go in line with larger sets of trajectories. Furthermore, periodic boundary conditions in combination with variable 
values for the used parameters $(\epsilon$, MinPts) will be probably necessary to advance to times at which the turbulent superstructures evolve gradually.
Parts of these efforts have been started already and will be presented elsewhere.

\acknowledgements
The work of CS and AP is supported by the Priority Programme on Turbulent Superstructures  of the Deutsche 
Forschungsgemeinschaft within Grant No. SPP 1881. JS wishes to thank the Tandon School of Engineering at 
New York University for financial support. We also acknowledge support with supercomputing resources 
by the John von Neumann Institute for Computing with project HIL12.

\end{document}